\documentclass[preprint,eqsecnum,prd,nofootinbib]{revtex4}
\def\Journal#1#2#3#4{{#1} {\bf #2}, #3 (#4)}

\def\NPA{{Nucl. Phys.} A}

\def\PLB{{Phys. Lett.}  B}
\def\PRL{Phys. Rev. Lett.}

\def\PRD{{Phys. Rev.} D}

\def\JHEP{JHEP}
\def\lsim{\mathrel{\rlap{\lower4pt\hbox{\hskip1pt$\sim$}}
    \raise1pt\hbox{$<$}}}                % less than or approx. symbol
\def\gsim{\mathrel{\rlap{\lower4pt\hbox{\hskip1pt$\sim$}}
    \raise1pt\hbox{$>$}}}                % greater than or approx. symbol
\def\msb{\overline{\rm MS}}              % MS bar

\def\bo{{\raise.15ex\hbox{\large$\Box$}}}         % D'Alembertian
\def\leftrightarrowfill{$\mathsurround=0pt \mathord\leftarrow \mkern-6mu
        \cleaders\hbox{$\mkern-2mu \mathord- \mkern-2mu$}\hfill
        \mkern-6mu \mathord\rightarrow$}       % <--> double differential
\def\dvec#1{\vbox{\ialign{##\crcr
        \leftrightarrowfill\crcr\noalign{\kern-1pt\nointerlineskip}
        $\hfil\displaystyle{#1}\hfil$\crcr}}}           % <--> accent
\def\be{\begin{equation}}
\def\ee{\end{equation}}

\def\bex{\begin{displaymath}}
\def\eex{\end{displaymath}}

\def\bea{\begin{eqnarray}}
\def\eea{\end{eqnarray}}

\def\beax{\begin{eqnarray*}}
\def\eeax{\end{eqnarray*}}

\usepackage{epsf}
\usepackage{graphicx}
\begin{document}

\title{The impact of new neutrino DIS and Drell-Yan data
\\ on large-$x$ parton distributions}

\author{J.F.~Owens$^a$, J.~Huston$^b$, C.E.~Keppel$^c$, S. Kuhlmann$^d$, 
J.G.~Morf\'{\i}n$^e$, F.~Olness$^f$, J.~Pumplin$^b$, and D.~Stump$^b$}

\vspace{.5 in}

\affiliation{$^a$Florida State University, $^b$Michigan State University,
$^c$Hampton University and Thomas Jefferson National Accelerator Facility, 
$^d$Argonne National Laboratory, 
$^e$Fermi National Laboratory,
$^f$Southern Methodist University}

\date{February 15, 2007}
\preprint{FSUHEP-070215}

\begin{abstract}
New data sets have recently become available for neutrino and antineutrino
deep inelastic scattering on nuclear targets and for inclusive dimuon
production in $pp {\rm \ and \ }pd$ interactions. These data sets are
sensitive to different combinations of parton distribution functions in the
large-$x$ region and, therefore, provide different constraints when
incorporated into global parton distribution function fits. We compare and
contrast the effects of these new data on parton distribution fits, with
special emphasis on the effects at large $x$. The effects of the use of
nuclear targets in the neutrino and antineutrino data sets are also
investigated.

\end{abstract}

\maketitle

\section{Introduction}
\label{sec:intro}

Global fits have long been used in the determination of parton distribution
functions (PDFs). By considering a variety of observables, one is sensitive
to different combinations of PDFs. In this way it has been
possible to place constraints on many of the different types of PDFs
in the nucleon. In particular, deep inelastic scattering with various
leptonic beams on nucleon targets and dilepton production in
hadron-hadron interactions have both played essential roles in the
determination of nucleon PDFs.

New data sets from the NuTeV \cite{nutev} and Chorus \cite{chorus}
Collaborations for neutrino and antineutrino interactions on iron
and lead targets, respectively, have recently become available. In
addition, the E866 Collaboration \cite{e866} has released new data sets
for inclusive dimuon production in $pp {\rm \ and \ } pd$ interactions.
These lepton-hadron and hadron-hadron cross sections are sensitive to
quite different combinations of PDFs and, therefore they will provide
significantly different types of constraints when incorporated into
a global fit.

It is the purpose of this paper to present the results of a series of
global fits that examine the impact of the new data sets when they
are included individually and in combination with a representative sample
of data for a wide variety of hard-scattering processes.

The use of nuclear targets for the neutrino and antineutrino data sets
complicates the extraction of nucleon PDFs. Model-dependent corrections
must be used to deduce the nucleon PDFs in such cases. The effects of
these nuclear corrections are examined in the course of performing the
various globals fits.

The outline of the paper is as follows. In the next section the various
theoretical tools used in the global fits are described along with
the models used for the nuclear target corrections. The methodology
used and the choices of data sets and kinematic cuts are also
presented. In Sec.~\ref{sec:fits} the details of the individual
fits are presented and the role of nuclear corrections is discussed in 
Sec.~\ref{sec:nuclear}. A strategy to reduce the model dependence for the 
$d/u$ ratio is discussed in Sec.~\ref{sec:DY} and Sec.~\ref{sec:conclusions} 
contains a summary of our conclusions.

\section{Theoretical and experimental input}
\label{sec:choices}

The primary goal of this analysis is to assess the impact of new DIS
neutrino and antineutrino data and new hadronic $\mu$ pair production data on
the PDFs in the large-$x$ region. Accordingly, it will be
useful to examine which combinations of PDFs these data sets are sensitive
to.

When using an isoscalar target, both the neutrino and
antineutrino charged current cross sections at large values of $x$ are 
proportional to the combination
\begin{equation}
\sigma^{\nu, \overline \nu} \propto u(x)+d(x)
\label{eqn:nudis}
\end{equation}
whereas muon or electron neutral current structure functions on proton
or deuteron targets are sensitive to the combinations
\begin{equation}
F_2^p \propto 4u(x)+d(x)
\label{eqn:pdis}
\end{equation}
and
\begin{equation}
F_2^d \propto u(x)+d(x).
\label{eqn:ddis}
\end{equation}
Thus, the heavy target charged current neutrino data and the deuterium
muon neutral current data are both sensitive to the same combination of
PDFs in the large-$x$ region.

This situation may be contrasted to that for dilepton production in
hadron-hadron collisions. The difference is most easily illustrated by
considering the kinematics for the production at a center-of-mass energy 
$\sqrt s$ of a lepton pair of mass $M$
and longitudinal momentum fraction $x_F$ using the lowest order process, for
which one has
\begin{equation}
x_{1,2}=\frac{\pm x_F + \sqrt{x_F^2 + 4M^2/s}}{2}.
\label{eqn:xf}
\end{equation}

For large values of $x_F$ the momentum fraction for the
beam parton $x_1$ approaches $x_F$ while that for the target 
parton $x_2$
approaches
$M^2/s$. For the kinematics of fixed-target experiments, one sees
that for large $x_F$ there is relatively little variation of $x_2$ and
it is typically rather small. Conversely, $x_1$ varies over a significant
range of values in the large-$x_F$ region. Therefore, one has approximately
\begin{equation}
\sigma^{pp} \propto 4u(x_1)\overline u(x_2) + d(x_1)\overline d(x_2)
\label{eqn:ppdy}
\end{equation}
whereas, due to the isoscalar nature of the deuteron target,
\begin{equation}
\sigma^{pd} \propto [4u(x_1)+d(x_1)][\overline u(x_2)+\overline d(x_2)].
\label{eqn:pddy}
\end{equation}
Thus, the large $x_F$ $pd$ lepton pair data are sensitive to the same
combination of large-$x$ PDFs as the proton target neutral current DIS 
cross section data.
Furthermore, the large-$x$ charged current neutrino data and the large $x_F$
lepton pair production data are sensitive to different combinations of PDFs.
This point will be important for understanding the results of the fits to be
presented in the next section.

All of the fits described in this analysis were done using
next-to-leading-order evolution in the $\msb$ scheme. The parametrization
chosen for the different flavors is identical to that used in the CTEQ6M
\cite{cteq6m} and CTEQ6.1M \cite{cteq61m} global fits. For data taken on
deuteron targets, nuclear corrections were included using a model-dependent 
estimate  
for the ratio $F_2^d/F_2^N$ from \cite{gomez} where $F_2^d {\ \rm and \ }
F_2^N$ are the structure functions for deuteron and isoscalar targets,
respectively. A simple parametrization was fitted to the extracted values of 
this ratio as given in Table X in \cite{gomez} and is included in the Appendix.
For the heavy targets used in the neutrino and antineutrino
data sets, the effects of nuclear corrections were included by using
the Kulagin-Petti parametrizations from Ref.~\cite{kp} for iron and lead 
targets in the case of
the NuTeV and Chorus data sets, respectively.

The basic impetus for this analysis originated with the observation that
the NuTeV data were expected to pull the valence distributions upward in the
region of large $x$ whereas
the E-866 data indicated that the valence distributions at large values of $x$
were already too high in the CTEQ6 PDFs. The NuTeV
data resulted from a follow-on experiment to that of the CCFR Collaboration
\cite{ccfr}. The results in the large-$x$ region are significantly
higher than those for CCFR in the $x$-bins at 0.55, 0.65, and 0.75. The
origin of this discrepancy is now understood by both groups and the NuTeV
data set is believed to be the more reliable of the two. Since the CCFR data
were included in the
global fits used to determine CTEQ6M and CTEQ6.1M PDFs,
replacing the CCFR data by those from NuTeV may well have a significant
impact on the PDF determination in the  large-$x$ region. However, the 
situation
is made more complex by the question of what nuclear corrections to use.
The previous CTEQ fits which used the CCFR data included nuclear
corrections that were based on measurements of $F_2$ in charged lepton
neutral current processes. The results of Ref.~\cite{kp} suggest that
the corrections for charged current neutrino interactions may differ from
these. Indeed, the issue of how large the nuclear corrections are at
large $x$ in neutrino interactions is model dependent at this time. Therefore,
it was decided to make a new reference fit starting with the CTEQ6.1M
PDFs and removing the CCFR data, while incorporating corrections for
deuteron targets where appropriate. This latter correction has not been
included in previous CTEQ fits. Furthermore, to better account for effects
due to heavy quarks, the ACOT scheme \cite{acot} has been implemented. This
primarily affects the low-$x$ region for neutrino induced processes and
so is somewhat secondary to the results to be presented below. In addition,
target mass corrections have also been included in all the fits.

The NuTeV data \cite{nutev} used in this analysis consist of measurements
of the cross sections for neutrino inclusive DIS (1170 data points)
and antineutrino inclusive DIS (966 data points). These data have
been corrected for electromagnetic radiative corrections 
using factors supplied by the NuTeV
collaboration. The data have seven sources of correlated systematic
errors in addition to a statistical error. The NuTeV collaboration
has published the correlated errors, which we use in our fitting
procedure in the manner described previously for the CTEQ6 PDFs 
\cite{cteq6m}. Similarly, the Chorus data \cite{chorus} consists 
of the neutrino and
antineutrino DIS cross sections (412 data points each) with 13
sources of experimental systematic error. These data have also been 
corrected for electromagnetic radiative effects.  

The Kulagin-Petti nuclear corrections include an isoscalar correction which 
accounts for the neutron excess in the corresponding Fe or Pb targets. For 
fits performed using these nuclear corrections the data sets without 
isoscalar corrections were used. For comparison purposes in later sections 
some results will be discussed which were obtained either without any nuclear 
corrections or with corrections which did not include the isoscalar 
correction. In these cases, the isoscalar corrections have been applied to 
the data using tables supplied by the NuTeV and Chorus Collaborations.

In addition, we fit the QCD theory to the cross
section data directly, unlike the previous CTEQ6 PDFs which were fit
to structure functions that had been extracted from the data by the
CCFR collaboration.

The E-866 data set consists of 184 (191) data points for the $pp (pd)$ 
cross sections $M^3 \frac{d^2\sigma}{dx_F \, dM}$. Correlated 
systematic errors are not available for these data, so the statistical and 
systematic errors have been added in quadrature.

\section{Global fits}
\label{sec:fits}

The reference fit was made to data from the BCDMS Collaboration \cite{bcdms}
for $F_2^p {\ \rm and \ } F_2^d$, from the NMC Collaboration \cite{nmc} for
$F_2^p {\ \rm and \ } F_2^d/F_2^p$, from H1~\cite{H1} and ZEUS~\cite{ZEUS} 
for $F_2^P$, from CDF \cite{cdf} and D\O\ \cite{d0}
for inclusive jet production, from CDF \cite{cdf_w} for the $W$
lepton asymmetry,
from E-866 \cite{e866_ratio} for the ratio of lepton pair cross sections for
$pd$ and $pp$ interactions, and from E-605 \cite{e605} for dimuon
production in $pN$ interactions. Note that the E-605 data were taken on
a copper target, but the nuclear corrections in the relevant kinematic region
have been measured \cite{dimuon_nuc} to be consistent with $A^1$, so no
nuclear corrections were included for these data.

\begin{figure}
\includegraphics[height=5 in]{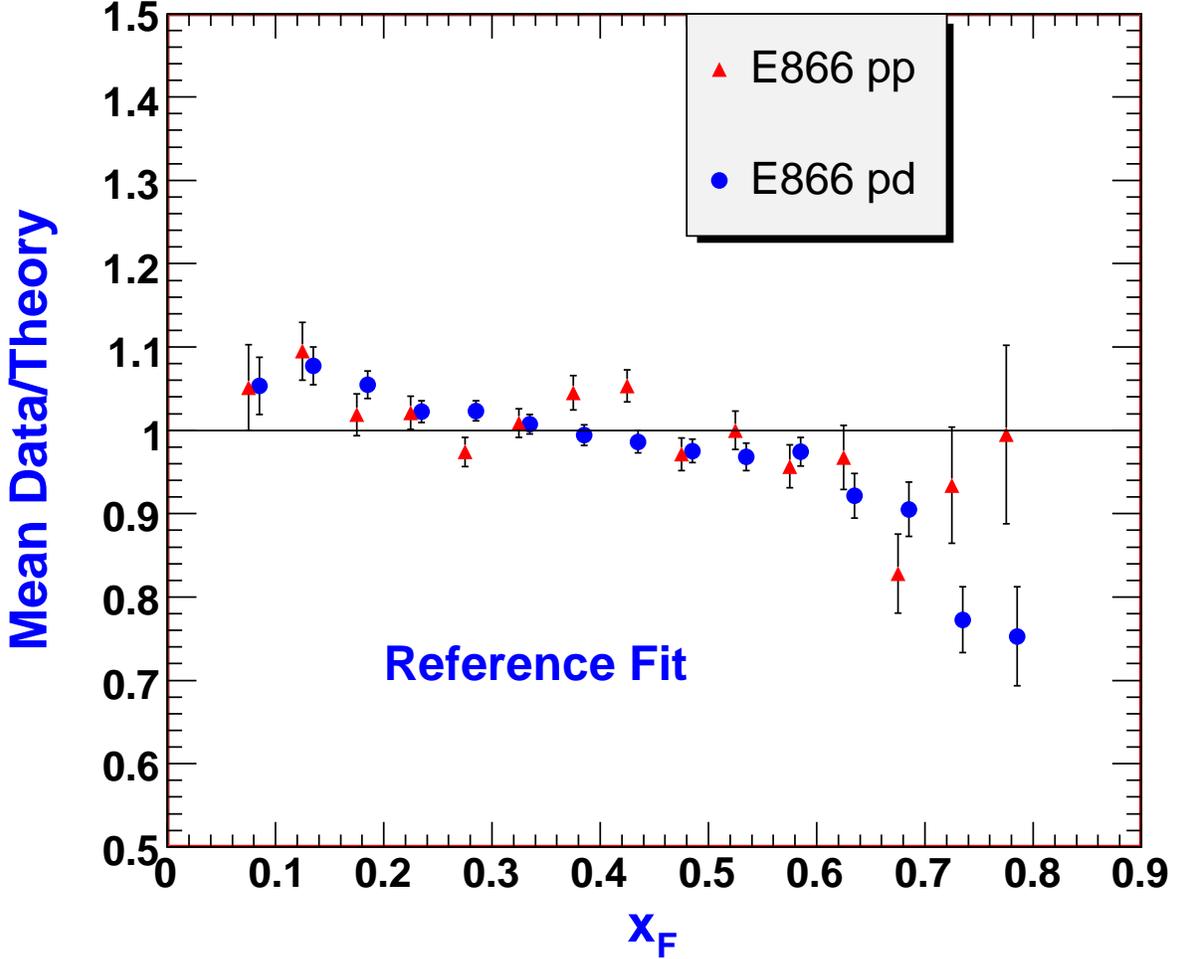}
\caption{Comparison between the reference fit and the E-866 data.}
\label{e866}
\end{figure}

In order to provide a context for the discussion of the various
global fits, Fig.~\ref{e866} shows a comparison between the
reference fit and the E-866 proton and deuterium data. The data are
plotted versus $x_F$ and integrated over the dimuon mass. What is
actually plotted in each $x_F$ bin is the weighted mean of
data/theory for that particular bin in an effort to display the two
dimensional data array in a simple fashion. Only the data
normalization has been fitted here; otherwise the reference fit
parameters are unchanged. The data,  both proton and deuterium,
are $\approx \! 5-10\%$ larger than theory at small $x_F$, then fall
below the theory at large $x_F$.  The largest $x_F$
deuterium data points appear to be lower than the proton data,
perhaps suggesting a nuclear correction for the deuteron is needed.  But
large $x_F$ corresponds to large $x_1$ ($\approx 0.8$ for the beam protons)
and small $x_2$ ($\approx 0.05$ for the target).  Based on the model used 
in this analysis, the deuterium corrections at these $x_2$ values are expected 
to be less than a few per cent and, therefore, insufficient to account for the 
difference between the theory and the $pd$ data.

\begin{figure}
\includegraphics[height=5 in]{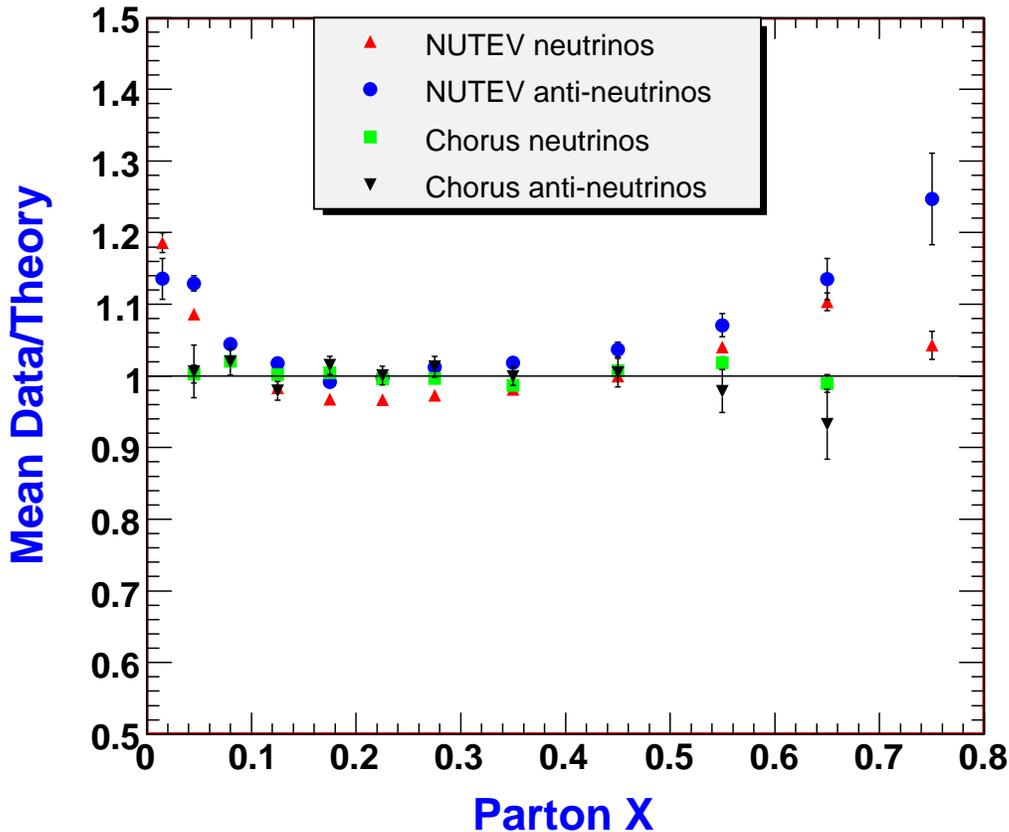}
\caption{Comparison between the reference fit and the Chorus and NuTeV
neutrino cross section data.}
\label{nutev_chorus}
\end{figure}

\begin{figure}
\includegraphics[height=5 in,angle=270]{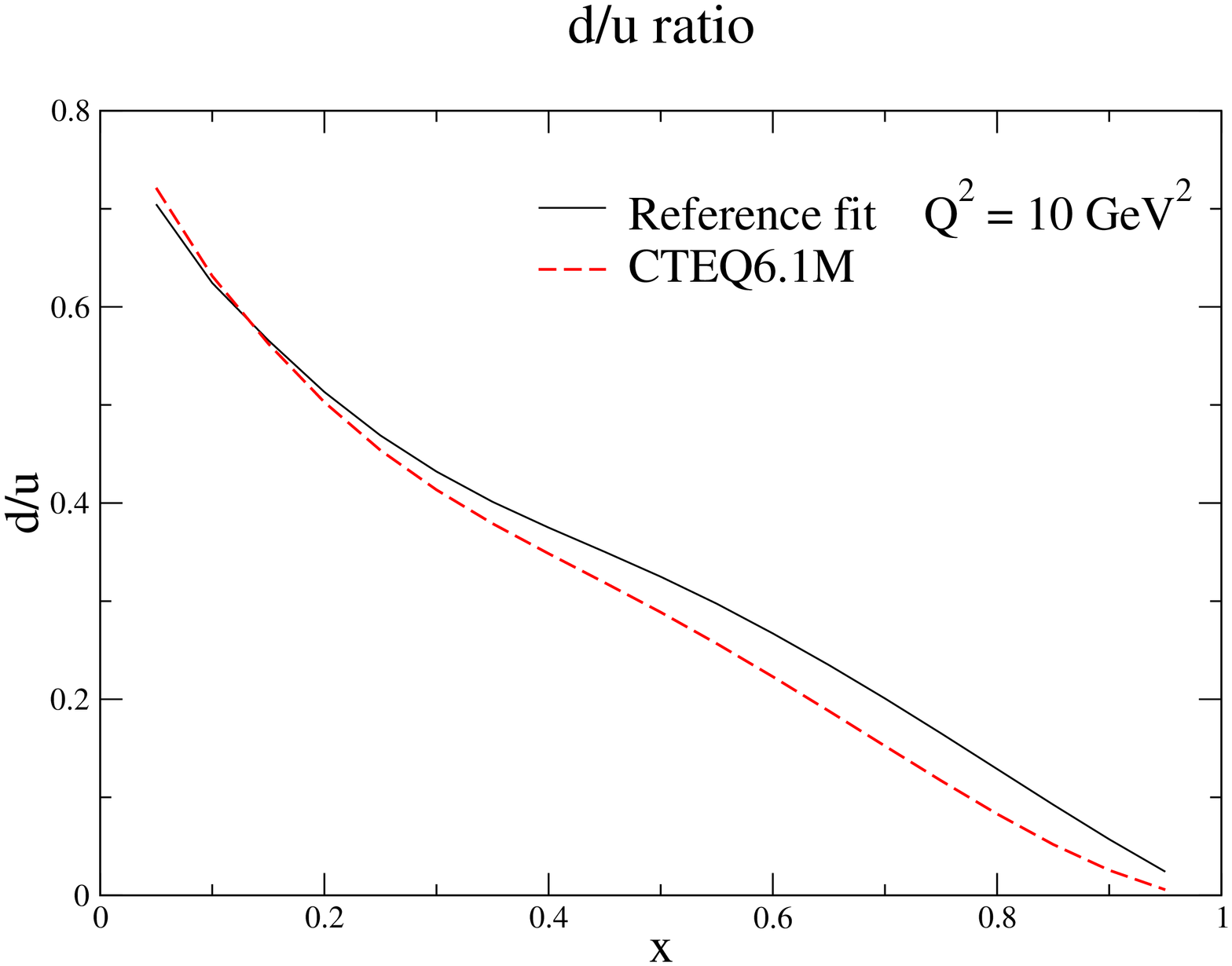}
\caption{Comparison of the $d/u$ ratios from the reference fit and
the CTEQ6.1M PDFs at $Q^2=10 {\rm \ GeV}^2$.}
\label{du_fig1}
\end{figure}

In a similar manner, Fig.~\ref{nutev_chorus} shows the NuTeV and Chorus
data sets plotted versus $x$ and integrated over $Q^2$. Again, in each $x$ bin
the weighted mean of data/theory is shown. In this case, as opposed to that
in Fig.~\ref{e866}, the NuTeV data at large values of $x$ are systematically 
higher
than the theory. It should be emphasized that these data have had nuclear
corrections applied to them. In the
large-$x$ region covered by the data in Fig.~\ref{nutev_chorus} the nuclear 
corrections  have the effect of raising the
data relative to the theory, which has
therefore made the agreement worse: the non-corrected data do not show
as significant an excess in this region. 
Another point to bear in mind is that the NuTeV and Chorus
collaborations have published both uncorrelated and correlated
errors. The correlated errors published by the experimental
collaborations are taken into account by the method that was
previously used for CTEQ6M \cite{cteq6m}. (This method is also used by
HERA analysis groups for electron and positron DIS.) For each source
of systematic error, a fitting parameter is introduced for systematic
variation of the data points within the published standard
deviations. We obtain the optimal values for these systematic shifts
to bring theory and data into closest agreement. The sizes of the
shifts are not unreasonably large---typically less than one standard
deviation. In Fig.~\ref{nutev_chorus}, the shifted data points are plotted in
comparison to the theory. The error bars then correspond only to the 
statistical errors.

The results shown in Figs.~\ref{e866} and \ref{nutev_chorus} illustrate how
the two types of data pull the valence distributions in opposite directions,
as discussed in the previous section.

For comparison purposes, the chi square values for the six data sets shown
in Figs.~\ref{e866} and \ref{nutev_chorus} are shown in the lower part of
Table I
in the column for the reference fit. Again, it should be emphasized that
these data sets were {\em not} included in the reference fit. The chi square
values are shown only so that one can gauge the changes that occur as the
various sets are added to the global fits.

As noted in Sec.~\ref{sec:choices}, the new data sets are primarily
sensitive at large values of $x$ to the PDF linear combinations $4 u + d$
and $u+d$. Accordingly, one might well expect the effects of adding
one or more of the new data sets to the reference fits to show up in the
$d/u$ ratio. Of course, there are data sets in the reference fit that 
are sensitive to these particular combinations already. Thus, if all
the data sets are consistent one would expect no change in the
$d/u$ ratio. On the other hand, should one or more data sets be inconsistent
with the others then either the $d/u$ ratio might change or the chi squares
for some or all of the data sets might increase, or both. Examples of each 
of those alternatives will be shown in the following discussion. 
These changes in the $d/u$ ratio will prove to be useful for characterizing 
the 
effects of adding or deleting data sets from the fits. It must be emphasized 
that these variations in the $d/u$ ratio results are not to be interpreted as 
``error bands.'' Furthermore, the variations resulting from different 
choices for data sets used in a given fit may well be smaller than the overall 
error bands estimated from the CTEQ6.1M 40 eigenvector uncertainty sets, for 
example. 
We are simply using the $d/u$ ratio results as a pedagogical tool to help 
understand the manner in which different data sets pull the results.  
 
In Fig.~\ref{du_fig1} the $d/u$ ratios are shown for both the CTEQ6.1M fit
and the reference fit described in the previous section. The inclusion of
deuteron corrections in the reference fit moves the data for the
$F_2^d/F_2^p$ ratio up in the region $0.4 \lsim x \lsim 0.8$ which is
reflected in the slight increase of the reference fit over the previous
CTEQ6.1M results. We also checked that the $d/u$ ratio in the reference fit 
was essentially unchanged from that in CTEQ6.1M if the deuteron corrections 
were not used, even though the CCFR data sets had been removed. An 
interesting footnote to this study is that equally good global fits can be 
obtained with or without the deuteron corrections. This point will be 
discussed further in Sec.~\ref{sec:DY}

In addition to the behavior of the $d/u$ ratio, the chi
square/point values for each data set and each fit are required in order to
understand the impact of adding each new data set. These values are shown in
Table~I.

\begin{figure}
\includegraphics[height=5 in,angle=270]{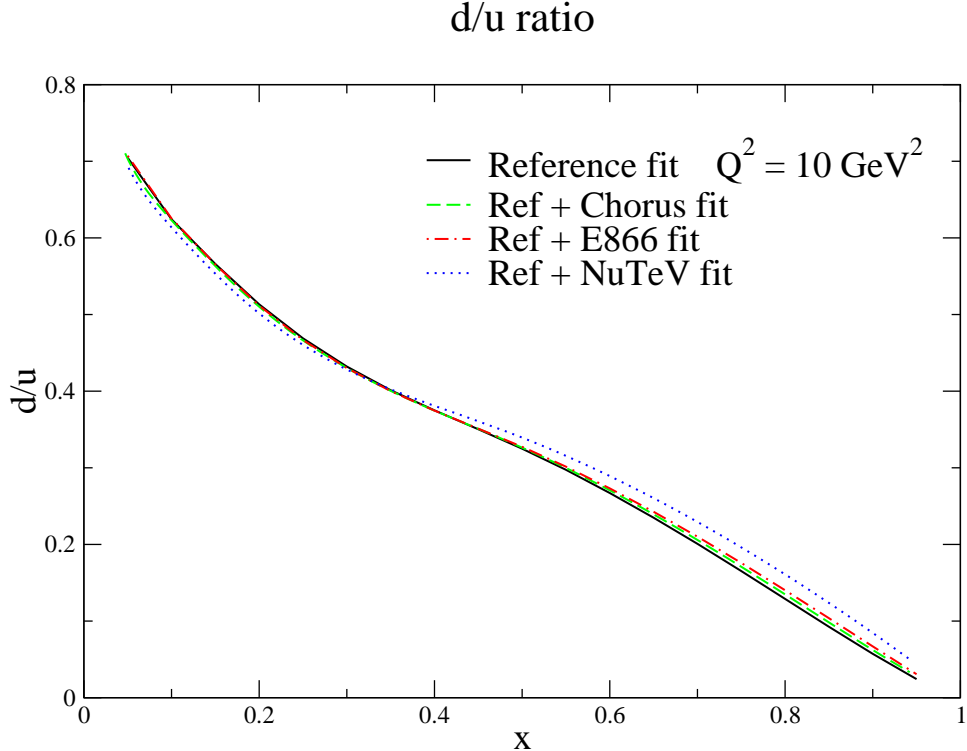}
\caption{Same as Fig.~\ref{du_fig1} with additional fits corresponding to
adding each new data set one at a time.}
\label{du_single}
\end{figure}

Next, the three data sets from Chorus, NuTeV, and E-866 are added to the
reference fit one at a time. The results are shown
in the fourth through sixth columns in Table~I and also in
Fig.~\ref{du_single}.
In each case, as one data set at a time is added, the results for the
$d/u$ ratio remain essentially unchanged from the reference fit. However,
examining the chi square values in Table~I shows that the NuTeV
data are forcing the chi square values higher for the BCDMS and NMC DIS data.
This is largely due to the increased values for the cross section in the
$x$ range above $x \approx .45$. Note that the relatively large number of
data points in the NuTeV sample allows it to dominate over the other DIS
data sets. Another possibility which will be discussed in detail later in this
section is that the large-$x$ nuclear corrections are too large. In this
kinematic region the ratio $\sigma^{\nu A}/\sigma^{\nu N}$ is less than one,
so the nuclear corrections raise the data when referenced back to an
isoscalar nucleon target. Reducing the nuclear corrections in this region
would
reduce the tension between the NuTeV data and those in the neutral
current DIS sets.

\begin{figure}
\includegraphics[height=5 in,angle=270]{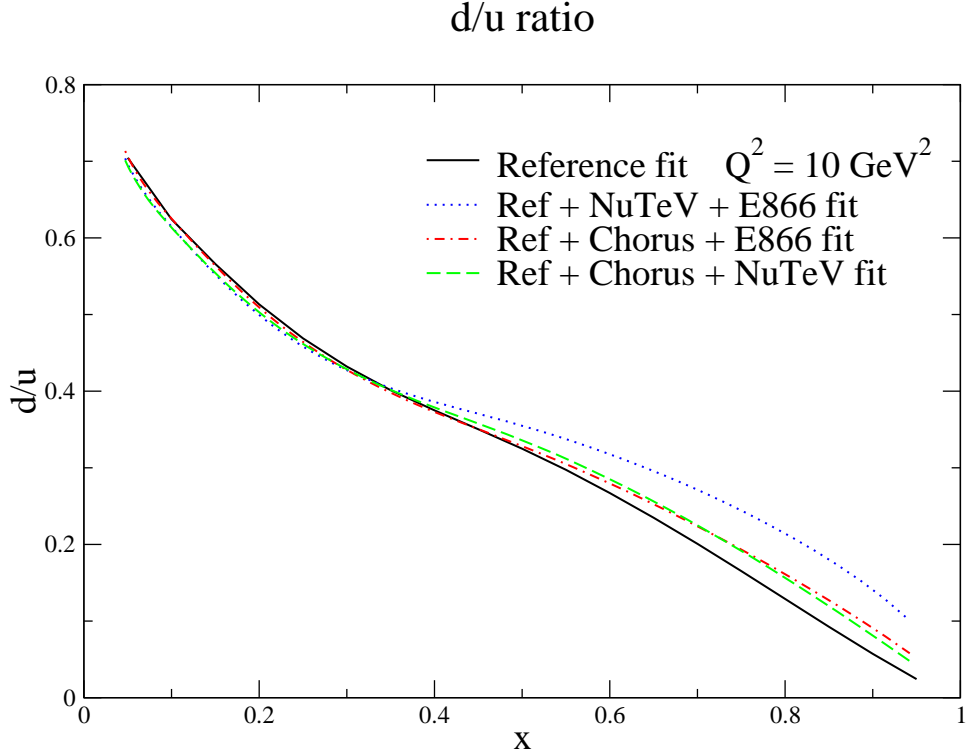}
\caption{Same as Fig.~\ref{du_fig1} with additional fits corresponding to
adding each new data set two at a time.}
\label{du_double}
\end{figure}

In the next step in the sequence of fits, the new data sets were included 
two at
a time. The results are shown in Fig.~\ref{du_double} and in columns
seven through nine in Table~I. The results for the Chorus+E866 and
Chorus+NuTeV fits are consistent with those obtained when the data sets were
added one at a time. However, a new result appears when the NuTeV and E-866
data sets are added at the same time. Fig.~\ref{du_double} shows that the
$d/u$ ratio at large values of $x$ has increased more than in the previous
fits. Furthermore, careful examination of the chi square values in Table~I
shows that the E-866 chi square has increased as well as that for the
NMC ratio data integrated over $Q^2$. This appears to be due to an interaction
between the conflicting demands of the NuTeV and E-866 data sets. On the one
hand, the NuTeV data require that the combination $u+d$ be increased at
large values of $x$. On the other hand, the best fits to the E-866 data
require that the combination $4u+d$ be reduced at large values of $x$. When
the fits are done with either data set alone added to the reference set it
is possible to get relatively good fits, although, as noted previously, the
NuTeV data do pull against the neutral current DIS data. When both the
NuTeV and E-866 data sets are added at the same time the fitting program 
finds a
new type of minimum with a significantly enhanced $d$ quark distribution in
the region of large $x$. In fact, the exponent of the $(1-x)$ factor
decreases from a typical value of 4.2 down to about 3.2, thereby flattening
out the $d$ quark distribution. This results in an increased $d/u$ ratio at
large values of $x$. This solution appears as a compromise --- increasing the
$d$ quark and slightly decreasing the $u$ quark allows an increase in
$u+d$ along with a decrease in $4u+d$. 
The price that is paid
is an increase in the NMC ratio data chi square. It should be emphasized
that this solution is a compromise between conflicting requirements: 
the resulting chi square values are generally not satisfactory.

\begin{figure}
\includegraphics[height=5 in,angle=270]{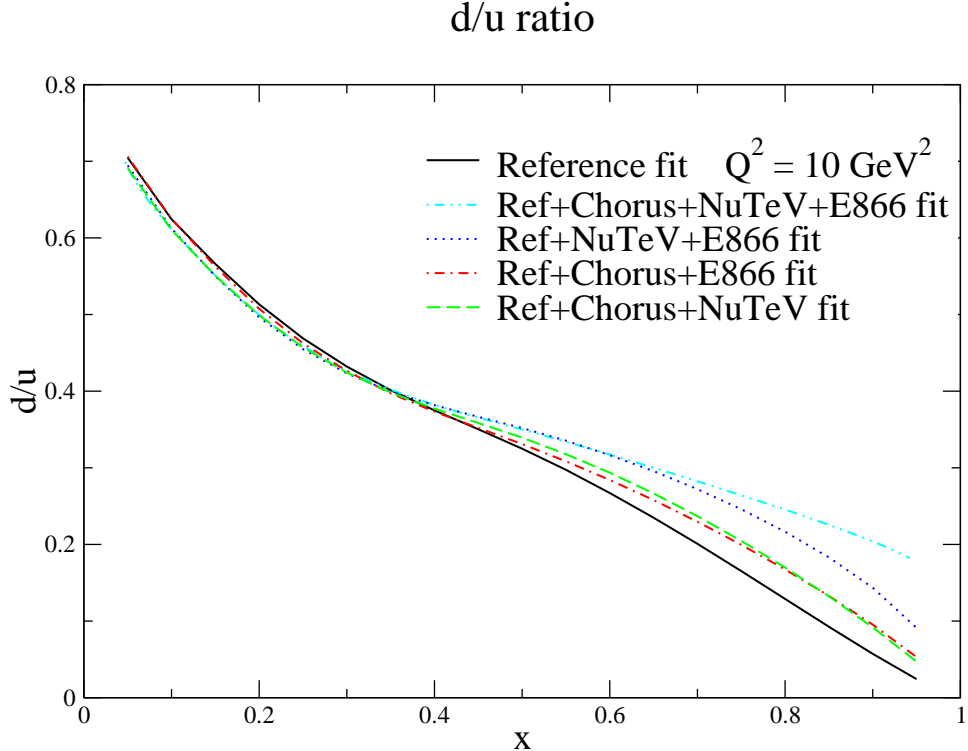}
\caption{Same as Fig.~\ref{du_double} with an additional fit corresponding to
adding all three new data sets.}
\label{du_triple}
\end{figure}

\begin{figure}
\includegraphics[height=5 in,angle=270]{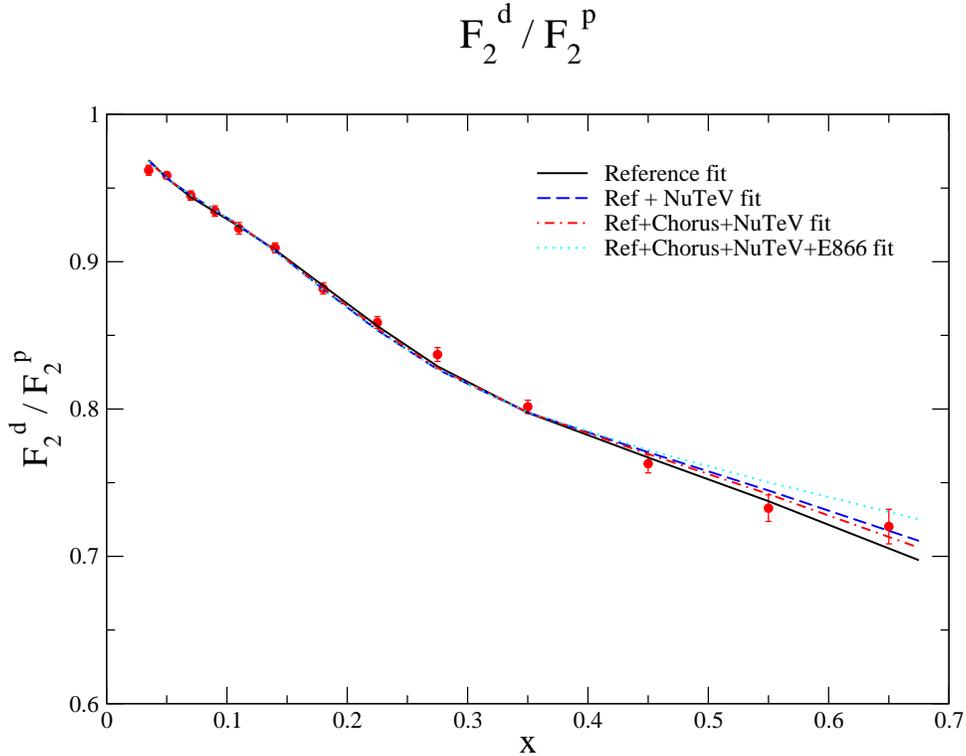}
\caption{Data for $F_2^d/F_2^p$ compared to the results of the fits
presented in the text.}
\label{f2df2p}
\end{figure}

The final step in the analysis is to add all three of the new data sets to
the reference set. The results are shown in Fig.~\ref{du_triple} and in the
column of Table~I labelled ``all.'' The same behavior as was seen in the case 
when both
the NuTeV and E-866 sets were added at the same time is seen here, as well.
There are elevated chi square values for the E-866 data, the NMC ratio data,
and for other neutral current DIS data sets. This again shows that the
NuTeV and E-866 data sets are pulling against each other.

The data for $F_2^d/F_2^p$ integrated over $Q^2$ \cite{nmc} have proven to be
instrumental in constraining the $d/u$ ratio in the fits discussed in the
text. These data are shown in Figure~\ref{f2df2p} along with several 
curves described in the text. The variation in $F_2^d/F_2^p$ is less than
that for the $d/u$ ratio since in the large-$x$ region if $d/u$ changes
from 0 to 1, $F_2^d/F_2^p$ only changes from 0.625 to 1. Nevertheless, the same
trend is apparent in curves as was shown previously for
the $d/u$ ratio.

The trends noted for the fits described above are not unique. This is a
result of the fact that the chi squares for some data sets increase
when other data sets are added to the fit. This tension between different
data sets can be explored by varying the weight for a given data set in the
total chi square. The NuTeV data tend to dominate due to the large number
of data points. Improved fits to the E-866 data or the NMC ratio data can
be obtained by simply multiplying their chi square contribution by a
weighting factor greater than one. If this is not done, then the smaller
data sets play little or no role in the chi square. Utilizing such
reweighting of individual chi squares results in families of fits rather
than one unique best fit.

\begin{table}
\begin{center}
\caption{Chi square/point values for each data set and each fit in this
analysis. Note: The last six data sets were not included in the
reference fit. Their chi squares were calculated after the fit and are
included to allow comparison with the values obtained in the subsequent fits.}
\footnotesize
\begin{tabular}{|c|c|c|c|c|c|c|c|c|c|c|c|}\hline
Data set & \# pts & Reference & Chorus & NuTeV & E-866 & Ch+866 & Nu+866 &
Ch+Nu & All & mod nuc & CTEQ6DU \\ \hline
BCDMS $F_2^p$ & 339 & 1.10 & 1.11 & 1.29 & 1.13 & 1.13 & 1.23 & 1.26 & 1.20
& 1.23 & 1.24 \\ \hline
BCDMS $F_2^d$ & 251 & 1.10 & 1.11 & 1.36 & 1.15 & 1.17 & 1.36 & 1.32 & 1.30
& 1.31 & 1.29 \\ \hline
H1 $F_2^p(1)$ & 79 & 0.93 & 0.94 & 1.20 & 0.94 & 0.93 & 1.11 & 1.22 & 1.12
& 1.03 & 1.03 \\ \hline
H1 $F_2^p(2)$ & 126 & 0.99 & 1.00 & 0.95 & 1.00 & 0.98 & 0.93 & 0.93 & 0.93
& 0.94 & 1.00 \\ \hline
H1 $F_2^p(3)$ & 130 & 0.77 & 0.78 & 0.76 & 0.76 & 0.76 & 0.74 & 0.75 & 0.73
& 0.73 & 0.76 \\ \hline
Zeus $F_2^p$ & 197 & 1.22 & 1.21 & 1.17 & 1.22 & 1.21 & 1.18 & 1.21 & 1.15
& 1.14 & 1.19 \\ \hline
NMC $F_2^d/F_2^p(x)$ & 13 & 1.05 & 1.01 & 1.20 & 1.11 & 1.02 & 1.60 & 1.06
& 1.35 & 1.20 & 0.85 \\ \hline
NMC $F_2^p$ & 201 & 1.46 & 1.47 & 1.79 & 1.47 & 1.47 & 1.78 & 1.81 & 1.77
& 1.64 & 1.56 \\ \hline
NMC $F_2^d/F_2^p(x,Q^2)$ & 123 & 0.96 & 0.96 & 0.99 & 0.97 & 0.97 & 1.11
& 1.03 & 1.09 & 0.97 & 0.92 \\ \hline
E-605 & 119 & 0.79 & 0.82 & 0.89 & 0.70 & 0.80 & 0.91 & 0.92 & 0.93 & 0.85
& 0.87 \\ \hline
CDF $W$asy & 11 & 1.15 & 1.08 & 1.08 & 1.26 & 1.15 & 1.23 & 1.14 & 1.25
& 1.02 & 1.28 \\ \hline
E866 $pd/2pp$ & 15 & 0.43 & 0.45 & 0.44 & 0.50 & 0.47 & 0.52 & 0.43 & 0.46
& 0.41 & 0.34 \\ \hline
D\O\  jets & 90 & 0.94 & 0.84 & 0.66 & 0.98 & 0.97 & 0.65 & 0.69 & 0.74
& 0.93 & 0.97 \\ \hline
CDF jets & 33 & 1.62 & 1.65 & 1.63 & 1.63 & 1.65 & 1.64 & 1.61 & 1.63
& 1.63 & 1.68 \\ \hline
Total chi square & - & 1947 \\ \cline{1-3}
\# points & - & 1727 \\ \hline \hline
E-866 $pp$ & 184 & {\it 1.23} & - & - & 1.16 & 1.15 & 1.22 & -
& 1.21 & 1.15 & 1.16 \\ \hline
E-866 $pd$ & 191 & {\it 1.85} & - & - & 1.49 & 1.49 & 1.84 & -
& 1.80 & 1.59 & 1.27 \\ \hline
NuTeV $\nu$ & 1170 & {\it 2.19} & - & 1.65 & - & - & 1.68 & 1.67
& 1.71 & 1.64 & - \\ \hline
NuTeV $\overline \nu$ & 966 & {\it 1.51} & - & 1.27 & - & - & 1.29
& 1.27 & 1.28 & 1.21 & - \\ \hline
Chorus $\nu$ & 412 & {\it 1.30} & 1.27 & - & - & 1.27 & - & 1.29
& 1.27 & 1.28 & - \\ \hline
Chorus $\overline \nu$ & 412 & {\it 1.08} & 1.09 & - & - & 1.08 & - & 1.16
& 1.15 & 1.18 & - \\ \hline
Total chi square & & 7453 & 2838 & 5218 & 2393 & 3357 & 5836 & 6247 & 6827
& 6606 & 2443 \\ \hline
\# points & - & 5062 & 2551 & 3863 & 2102 & 2926 & 4238 & 4687 & 5062 & 5062
& 2102 \\ \hline
\end{tabular}
\end{center}
\end{table}

\section{Nuclear Corrections}
\label{sec:nuclear}

\begin{figure}
\includegraphics[height=5 in]{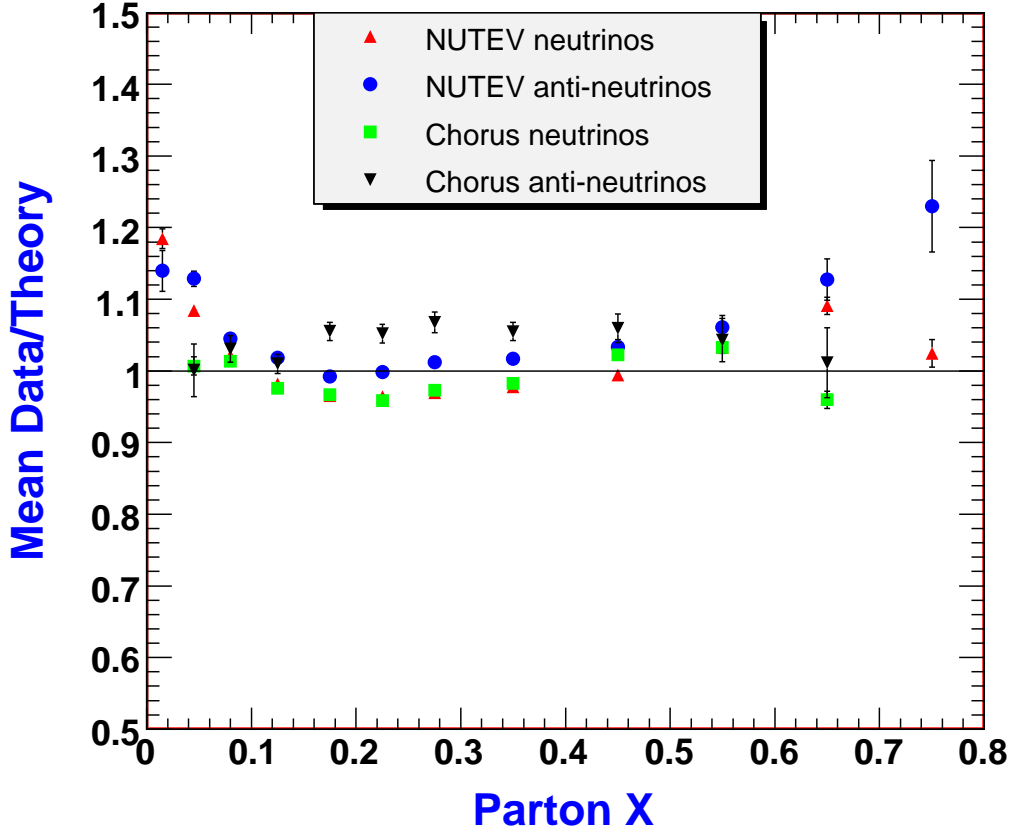}
\caption{Comparison between the reference fit and the unshifted Chorus and
NuTeV neutrino data. Kulagin-Petti nuclear corrections are included.}
\label{no_shift}
\end{figure}

A review of the chi square values in Table I shows that the Chorus data
are reasonably well described by the reference fit and by all of the other
fits as well. This situation is in contrast to that for the NuTeV data,
where the chi squares are significantly higher. This is shown, for example,
in Fig. \ref{nutev_chorus}. In this figure the Kulagin-Petti nuclear
corrections for Fe and Pb have been applied to  the NuTeV and Chorus 
experiments,
respectively. In order to fully understand the significance of this figure,
one must be aware that the data have been shifted by the optimal values
of the correlated errors prior to making the plot. It turns out that the
correlated errors allow larger shifts for the Chorus data than for the NuTeV 
data. Thus,
there is more freedom for the Chorus data to shift and to thereby reduce
the chi square. This is shown in more detail in Fig.~\ref{no_shift} where
now the {\em unshifted} data are plotted.  The deviation of the
mean data/theory values from one is now more pronounced in some regions of 
$x$ for the Chorus
data than for the NuTeV data. Comparison with Fig.~\ref{nutev_chorus} shows
that the NuTeV data have been shifted very little while the Chorus shifts are
much larger. Hence, in the various fits discussed previously it is clear that
much more weight is being given to the NuTeV data due to their smaller 
systematic errors.
Thus, it is premature to conclude that there is any systematic disagreement
between the two experiments simply because of the chi square values shown
in Table I.

\begin{figure}
\includegraphics[height=5 in, angle = 270]{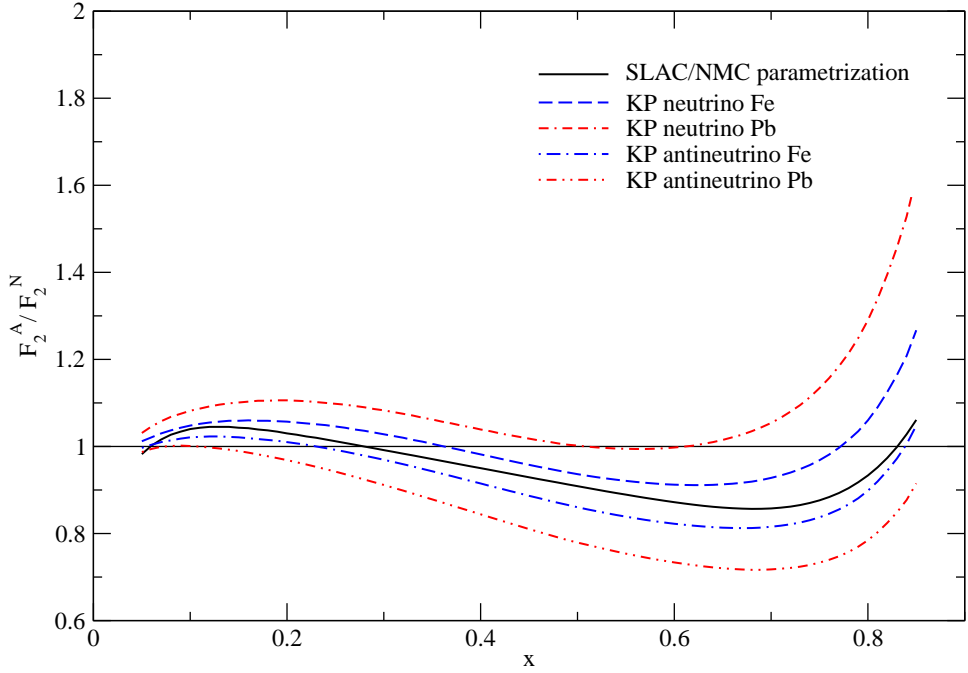}
\caption{Comparison of various nuclear corrections used in this analysis 
expressed as ratios 
of $F_2$ on a nucleus $A$ to that for an isoscalar nucleon target.}
\label{nuk}
\end{figure}

\begin{figure}
\includegraphics[height=5 in]{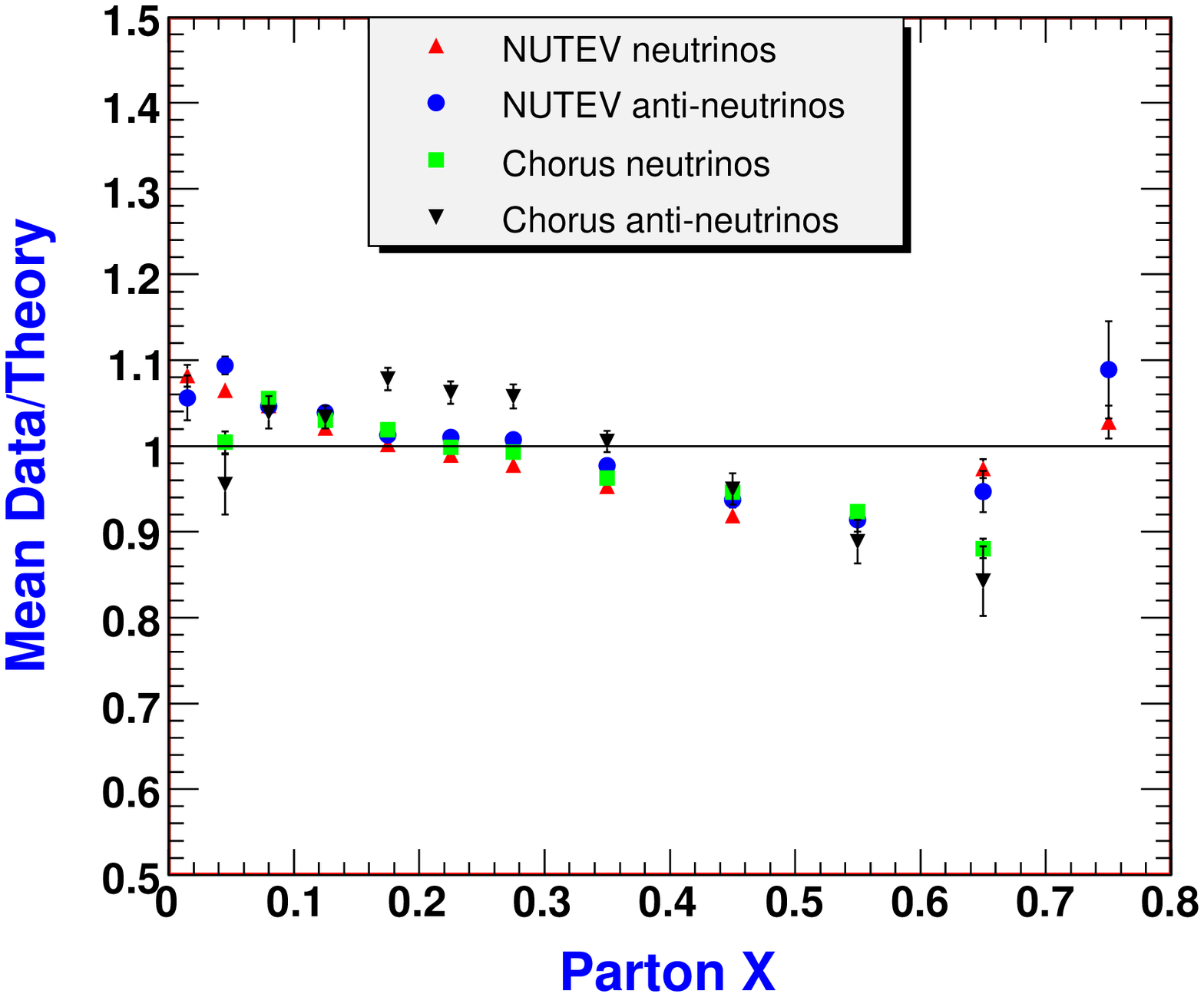}
\caption{Comparison between the reference fit and the unshifted Chorus and
NuTeV neutrino data without any nuclear corrections.}
\label{no_shift_no_cor}
\end{figure}

It is instructive to examine the typical magnitude of the nuclear corrections 
used in this analysis. These are shown in Fig.~\ref{nuk} for the structure 
function $F_2$ in the form of a ratio of the nuclear structure function to 
that of an isoscalar nucleon target. The SLAC/NMC curve is the result of an 
A-independent parametrization fit to calcium and iron charged-lepton 
DIS data\cite{SLAC/NMC}. The other four curves show the Kulagin-Petti 
results at $Q^2=20 
{\ \rm GeV}^2$ . The results for $F_1 {\ \rm and \ } xF_3$ are similar. 
Note, in 
particular, that the corrections for antineutrinos are larger than for charged 
leptons while those for neutrinos are generally smaller.

This issue may be investigated in a different fashion.
Fig.~\ref{no_shift_no_cor} shows the results of removing the nuclear
corrections used in Fig.~\ref{no_shift} while still using {\em unshifted}
data. Now there is rather good agreement between the different data sets in
the sense that the mean data/theory values are consistent even though they
are not equal to one. Moreover, the pattern of deviation from unity in the
large-$x$ region does look very much like the usual pattern associated with
nuclear corrections. Thus, these results suggest on a purely
phenomenological level that the nuclear corrections may well be very
similar for the $\nu$ and $\overline \nu$ cross sections and that the
overall magnitude of the corrections may well be smaller than in the
model used in this analysis.

In order to test this idea, several fits were done using nuclear
corrections determined experimentally in charged-lepton DIS
experiments (The SLAC/NMC curve in Fig.~\ref{nuk}) but with the magnitude 
reduced by a scale factor $\eta$, 
{\it i.e.,} the nuclear correction ratio $R$ was replaced by $1-\eta (1-R)$ 
so that $\eta=0$ gives no correction and $\eta=1$ gives the full correction. 
The same corrections were used for both the $\nu$ and $\overline \nu$
cross sections for both experiments. The chi squares for the 
NuTeV data were
significantly reduced when the nuclear corrections were reduced by a
factor on the order of $0.3 \pm 0.3$. The results for $\eta=0.3$ are shown 
in Table~I in the column labelled ``mod nuc.''  
As expected, the chi squares
for the Chorus data were essentially unchanged, since the data could
simply shift to accommodate the nuclear correction model. While
certainly not definitive, this observation strongly suggests that one
must have a better understanding of the nuclear corrections in
neutrino interactions before the NuTeV data can be understood in the
context of a global fit. Regarding the results for the $d/u$ ratio,
it may well be that the least model-dependent determination should
be made using data not taken on heavy targets. In this sense then,
fits such as the Reference fit together with the E-866 data should
give a less model-dependent result. Such fits are the focus of the 
next Section.

One last point concerning the nuclear corrections should be emphasized. The 
Kulagin-Petti corrections include a correction for the neutron excess in 
the relevant nucleus. This is why the correction factors for Fe and Pb are so 
different in Fig.~\ref{nuk}. Thus, when these nuclear corrections are used 
one must use data that have not been corrected for the neutron excess. 
On the other hand, when comparing the data to theory without the nuclear 
corrections one must be sure to include a correction for the neutron excess 
either in the theory or in the data.

\section{Determination of $d/u$ without heavy target data}
\label{sec:DY}

\begin{figure}
\includegraphics[height=5 in]{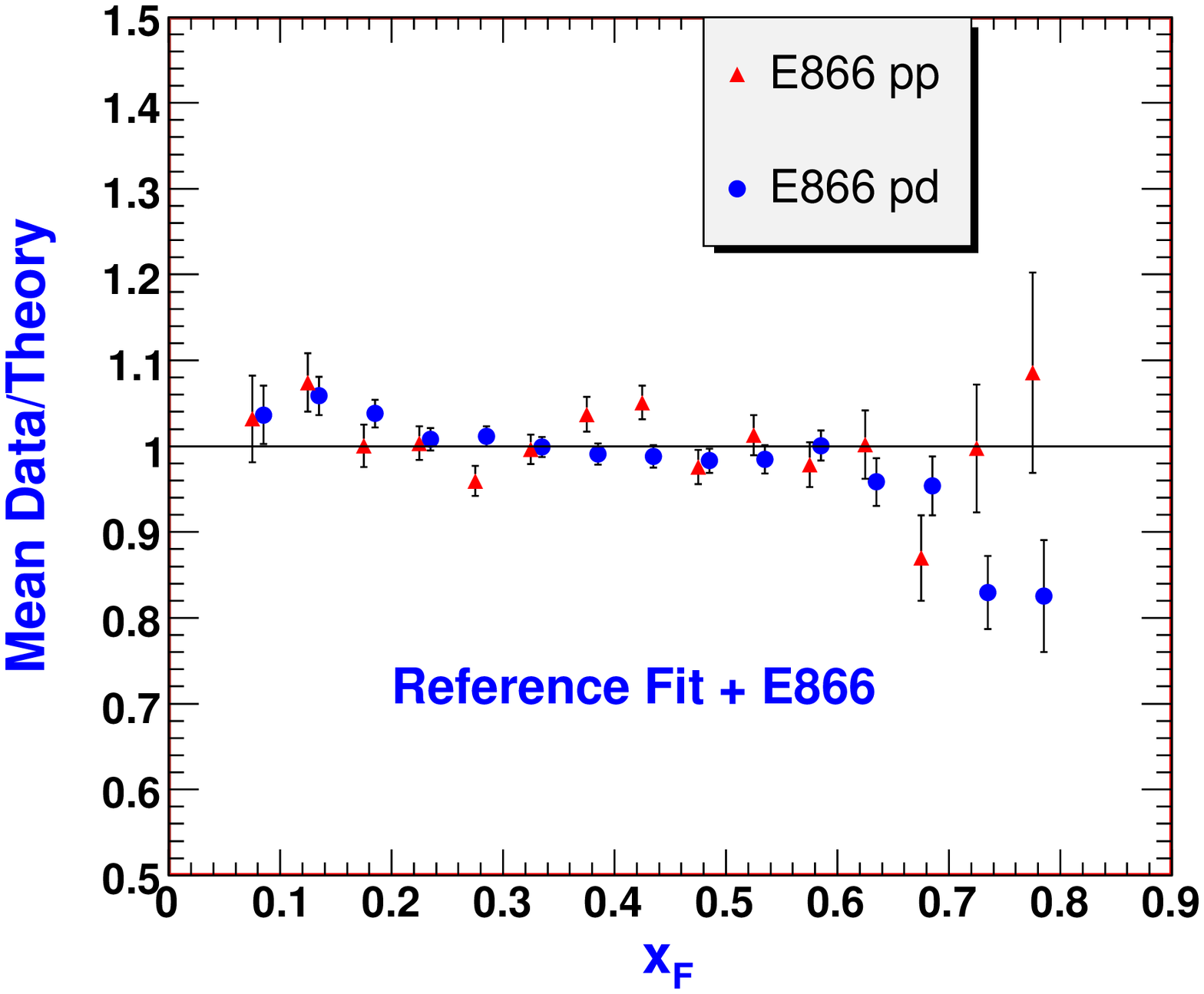}
\caption{Comparison of the global fit including E-866, with the
E-866 data.} \label{ref_e866}
\end{figure}

\begin{figure}
\includegraphics[height=5 in]{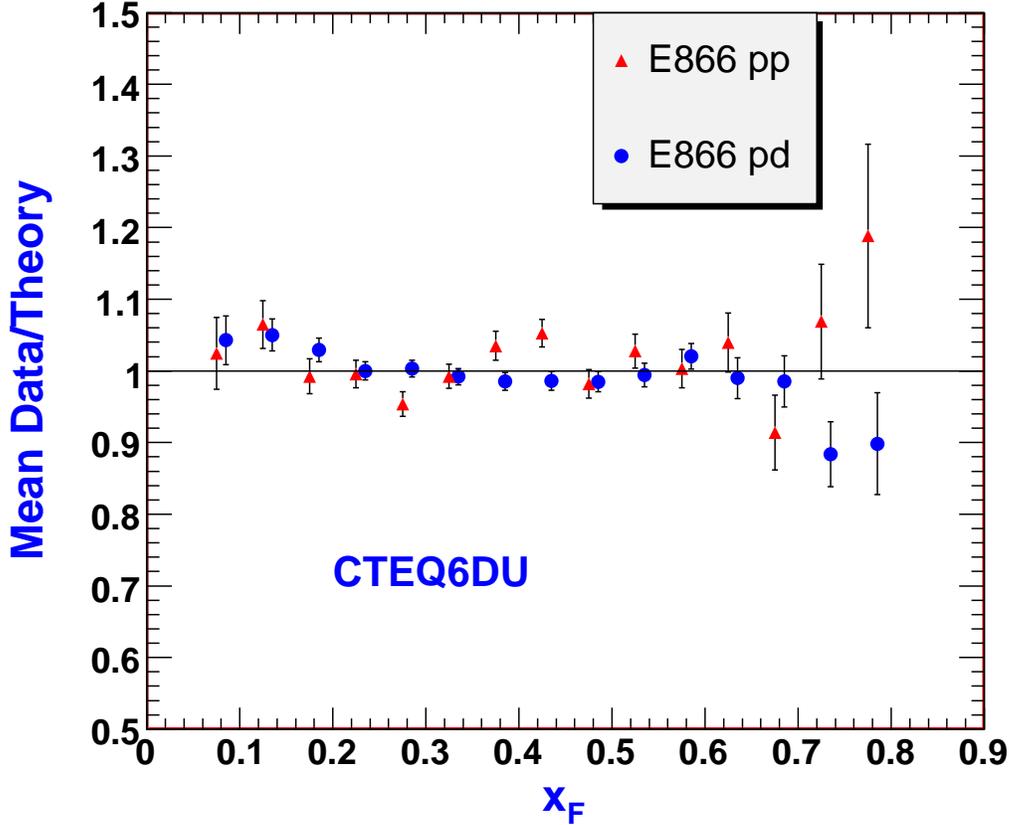}
\caption{Comparison of the global fit emphasizing the fixed target
and collider Drell-Yan data,  with the E-866 data.}
\label{force_e866}
\end{figure}

\begin{figure}
\includegraphics[height=5 in]{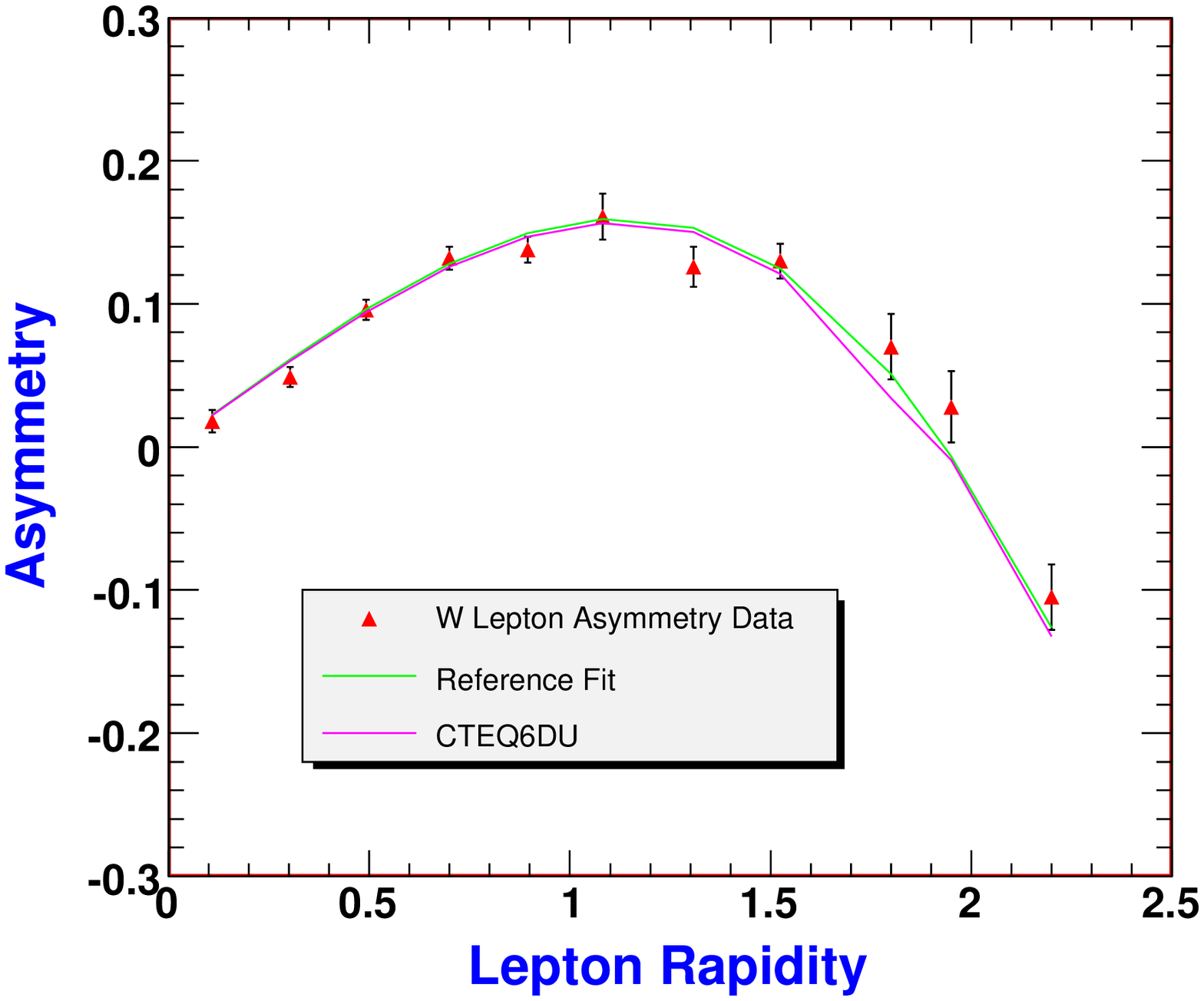}
\caption{Comparison of the global fit emphasizing the fixed target
and collider Drell-Yan data, with the CDF W lepton asymmetry data.}
\label{wasy}
\end{figure}

In this section we focus on the E-866 data 
and the question of whether they can be well described in the context of 
a global fit without using data involving heavy targets.
  
As noted earlier, at large values of $x_F$ these data help determine the 
$d/u$ ratio. For $x<0.3$, however,
the CDF W lepton asymmetry data and the $F_2^d/F_2^p$ ratio data
from NMC provide strong constraints that must be satisfied to be
sure we have a consistent global picture.

Fig.~\ref{ref_e866} shows the starting point,  the Reference fit +
E-866 mentioned earlier.  The $\chi^2$/data point is 1.16 for the $pp$
data and 1.49 for the $pd$ data (which has better statistics). These 
values are  much
improved from the Reference fit without E-866 data but are still not
acceptable. The differences between the $pp$ and $pd$ data at large
$x_F$ were discussed earlier, along with the reasons
we believe that this is not due to deuterium effects but is more
likely a statistical fluctuation.

In an effort to improve the description of the E-866 data sets, a fit was 
performed in which 
additional weight was given to E-866 as well as the W asymmetry and
$F_2^d/F_2^p$ data. This fit is referred to as CTEQ6DU and the corresponding 
chi square values are shown in the last column of Table~I. 
The $\chi^2$/data point is 1.16 for the $pp$ data
and 1.27 for the $pd$ data, the latter value showing a significant 
improvement.  In addition,  it appears
this fit provides a compromise solution between the $pp$ and $pd$ data
points at large $x_F$.

This fit provides a good overall description of the data. The total
$\chi^2$ is 2443 for 2102 data points, which is slightly higher than the
Reference fit + E-866 without the extra statistical weight (which
had a total $\chi^2$ of 2393 as shown in column 6 of Table I).  
The CDF W lepton asymmetry data in
this fit has a $\chi^2$/point of 1.28, and is shown in
fig.~\ref{wasy}.  This figure shows that the change in the theoretical
prediction from the original reference fit is very small. The 
$F_2^d/F_2^p$ ratio data from NMC,  shown earlier,  has a
$\chi^2$/point of 0.85.  Finally, the $d/u$ ratio resulting from this fit 
is essentially the same as that for the Reference fit shown earlier.

One final comment concerns the dependence of the global fits on the 
model dependent deuteron corrections. The CTEQ6DU fit (with the extra 
weighting discussed above) was repeated, but without the deuteron corrections 
for the DIS experiments with deuteron targets. The resulting chi square was 
essentially unchanged from the CTEQ6DU fit (2444 instead of 2443) and there 
were very minor changes amongst the various experiments. The resulting 
$d/u$ ratio was essentially the same as for CTEQ6.1M. Thus, equally good fits 
could be obtained with or without the deuteron corrections and the only 
visible result was the change in the $d/u$ ratio as shown in 
Fig.~\ref{du_fig1}. So, on the basis of these fits alone, one could not 
determine the necessity of including the deuteron corrections. This point 
will be explored further in a future analysis.

\section{Conclusions}
\label{sec:conclusions}

The impact of the new data sets from the NuTeV~\cite{nutev},
Chorus~\cite{chorus}, and E-866~\cite{e866} Collaborations on the
behavior of PDFs at large values of $x$ has been studied. The
effects are most pronounced when one examines the $d/u$ ratio. Adding each
new data set one at a time causes no significant alterations to the standard
reference fit, which is based on the CTEQ6.1M~\cite{cteq61m} PDFs. Adding
Chorus and E-866 or adding Chorus and NuTeV likewise causes little change.
However, adding NuTeV and E-866 simultaneously causes the $d/u$ ratio to
flatten out substantially, as the program is able to find a new minimum
which represents a compromise between the conflicting demands of the
two data sets at large $x$. This effect is enhanced further when all three
data sets are added to the reference set.

One conclusion is that the NuTeV data set together with the model used
for the nuclear corrections pulls against several of the other data
sets, notably the E-866 and the BCDMS and NMC sets. Reducing the nuclear
corrections at large values of $x$ would lessen the severity of this pull
and would result in improved chi square values.

\vspace{.25in}
\noindent\textbf{Acknowledgement}

The authors wish to thank Wu-Ki Tung for his interest in this topic and for 
discussions about the relative impact of the different data sets on global 
fits. This work was supported in part by the National Science Foundation and 
the U.S. Department of Energy.

\section{Appendix}
\label{sec:appendix}

The deuteron correction factor which is used to convert deuteron target data
to that corresponding to an isoscalar target was obtained by fitting the
model-dependent results for $F_2^d/F_2^N$ in Table X of Ref.~\cite{gomez}. 
The parametrization
used is
$$ F_2^d/F_2^N=\sum_{j=1}^6 a_j x^{j-1}/(1-\exp^{-a_7(1/x-1)}).$$

One can then divide the deuteron data by the above parametrization to obtain
data corresponding to an isoscalar target. The parameter values are given in
Table II.

\begin{table}[h]
\begin{center}
\caption{Parameter values for the deuteron correction parametrization.}
\begin{tabular}{|c|r|}\hline
Parameter & Value \\ \hline
$a_1$ &  1.0164  \\ \hline
$a_2$ & -0.0478  \\ \hline
$a_3$ & -0.1335  \\ \hline
$a_4$ &  0.3530  \\ \hline
$a_5$ &  0.2272  \\ \hline
$a_6$ & -1.2906  \\ \hline
$a_7$ &  5.6075  \\ \hline
\end{tabular}
\end{center}
\end{table}

\vfill \eject


\begin{thebibliography}{99}

%NuTeV data
\bibitem{nutev} M. Tzanov et al., \Journal{PRD}{74}{012008}{2006}

%Chorus data
\bibitem{chorus}
G. Onengut et al., \Journal{\PLB}{632}{65}{2006}

%E866 data
\bibitem{e866} Jason Webb, Ph.D. Thesis, New Mexico State University, 2002,
arxiv:hep-ex/0301031 and Paul Reimer, Private Communication.

%CTEQ6M
\bibitem{cteq6m} J. Pumplin et. al., \Journal{\JHEP}{0207}{012}{2002}

%CTEQ6.1M
\bibitem{cteq61m} D. Stump, et al., \Journal{\JHEP}{0310}{046}{2003}

%Gomez et al deuteron corrections
\bibitem{gomez} J. Gomez et al., \Journal{\PRD}{49}{4348}{1994}.

%Kulagin-Petti nuclear corrections
\bibitem{kp} S.A. Kulagin and R. Petti, \Journal{\NPA}{765}{126}{2006}

%CCFR
\bibitem{ccfr}
CCFR Collaboration: W.\ G.\ Seligman {\it et al.},
%``Improved Determination Of Alpha(S) From Neutrino Nucleon Scattering,''
Phys.\ Rev.\ Lett.\ {\bf 79} (1997) 1213 [hep-ex/970107];
CCFR Collaboration: U.\ K.\ Yang {\it et al.},
Phys.\ Rev.\ Lett.\ {\bf 86} (2001) 2742 [hep-ex/0009041].


%ACOT
\bibitem{acot}
M.\ A.\ Aivazis, F.\ I.\ Olness and W.\ K.\ Tung,
%``Leptoproduction of heavy quarks. 1. General formalism and kinematics of charged current and neutral current production processes,''
Phys.\ Rev.\ {\bf D 50} (1994) 3085 [hep-ph/9312318];
%%CITATION = HEP-PH 9312318;%%
M.\ A.\ Aivazis, J.\ C.\ Collins, F.\ I.\ Olness and W.\ K.\ Tung,
%``Leptoproduction of heavy quarks. 2. A Unified QCD formulation of charged and
neutral current processes from fixed target to collider energies,''
Phys.\ Rev.\ {\bf D 50} (1994) 3102 [hep-ph/9312319].
%%CITATION = HEP-PH 9312319;%%

%BCDMS
\bibitem{bcdms}
BCDMS Collaboration: A.\ C.\ Benvenuti {\it et al.},
%``A High Statistics Measurement Of The Proton Structure Functions F(2) (X, Q**2) And R From Deep Inelastic Muon Scattering At High Q**2,''
Phys.\ Lett.\ {\bf B 223} (1989) 485;
BCDMS Collaboration: A.\ C.\ Benvenuti {\it et al.},
Phys.\ Lett.\ {\bf B 236} (1989) 592.

%NMC
\bibitem{nmc}
New Muon Collaboration: M.\ Arneodo {\it et al.},
%``Measurement of the proton and deuteron structure functions, F2(p) and  F2(d), and of the ratio sigma(L)/sigma(T),''
Nucl.\ Phys.\ {\bf B 483} (1997) 3 [hep-ph/9610231]; and
M.~Arneodo {\it et al.},
%``Accurate measurement of F2(d)/F2(p) and R(d)-R(p),''
Nucl.\ Phys.\ B {\bf 487} (1997) 3
[hep-ex/9611022].

%H1
\bibitem{H1}
H1 Collaboration: C.\ Adloff {\it et al.},
Eur.\ Phys.\ J.\ {\bf C 13} (2000) 609 [hep-ex/9908059]; %
%``Measurement of neutral and charged current cross sections in electron  proton collisions at high Q**2,''
Eur.\ Phys.\ J.\ {\bf C 19} (2001) 269 [hep-ex/0012052]; %
%%CITATION = HEP-EX 0012052;%%
Eur.\ Phys.\ J.\ {\bf C 21} (2001) 33 [hep-ex/0012053].

%ZEUS
\bibitem{ZEUS}
ZEUS Collaboration: S. Chekanov {\it et al.},
Eur.\ Phys.\  J.\ {\bf C 21} (2001) 443 [hep-ex/0105090];
A.M. Cooper-Sarkar,
Proceedings of International Europhysics Conference
on HEP 2001, Budapest [hep-ph/0110386].


%CDF jets
\bibitem{cdf}
CDF Collaboration: T. Affolder {\it et al.},
Phys.\ Rev.\ {\bf D 64} (2001) 032001 [hep-ph/0102074].

%D0 jets
\bibitem{d0}
D\O\ Collaboration: B. Abbott {\it et al.},
Phys.\ Rev.\ Lett.\ {\bf 86} (2001) 1707 [hep-ex/0011036];
and Phys.\ Rev.\ {\bf D 64} (2001) 032003 [hep-ex/0012046].


%CDF W asymmetry
\bibitem{cdf_w}
CDF Collaboration: F.\ Abe {\it et al.},
%``Measurement of the lepton charge asymmetry in W boson decays produced  in 
%p anti-p collisions,''
Phys.\ Rev.\ Lett.\ {\bf 81} (1998) 5754 [hep-ex/9809001].

%E-866 ratio
\bibitem{e866_ratio}
E866 Collaboration: R.\ S.\ Towell {\it et al.},
Phys.\ Rev.\ {\bf D 64} (2001) 052002 [hep-ex/0103030].

%E-605
\bibitem{e605}
E605 Collaboration: G.\ Moreno {\it et al.},
Phys.\ Rev.\ {\bf D 43} (1991) 2815.

%SLAC/NMC routine
\bibitem{SLAC/NMC}
A. Bruell, private communication.

%dimuon nuclear efects
\bibitem{dimuon_nuc} D.M. Alde et al., \Journal{\PRL}{64}{2479}{1990}.

%SMU nuclear corrections ref.
%\bibitem{smu} SMU nuclear corrections study

\end{thebibliography}
\end{document}